# The Sloan Digital Sky Survey Science Archive

## Migrating a Multi-Terabyte Astronomical Archive from Object to Relational DBMS

**Aniruddha R. Thakar**
**Alexander S. Szalay**
The Johns Hopkins University

**Peter Z. Kunszt**
CERN

**Jim Gray**
Microsoft Research

**Version #3**

**March 12, 2004**

# Abstract

The Sloan Digital Sky Survey Science Archive is the first in a series of multi-Terabyte digital archives in Astronomy and other data-intensive sciences. To facilitate data mining in the SDSS archive, we adapted a commercial database engine and built specialized tools on top of it. Originally we chose an object-oriented database management system due to its data organization capabilities, platform independence, query performance and conceptual fit to the data. However, after using the object database for the first couple of years of the project, it soon began to fall short in terms of its query support and data mining performance. This was as much due to the inability of the database vendor to respond our demands for features and bug fixes as it was due to their failure to keep up with the rapid improvements in hardware performance, particularly faster RAID disk systems. In the end, we were forced to abandon the object database and migrate our data to a relational database. We describe below the technical issues that we faced with the object database and how and why we migrated to relational technology.

# 1 Introduction

The advent of digital archives, enabled by quantum leaps in the technology to publish, distribute and mine data over the Internet, has given rise to a data avalanche in many branches of science and engineering. The Human Genome Project and the Large Hadron Collider are two examples of very large scientific datasets coming online in the biological and particle physics communities respectively. Astronomy is no exception, and is perhaps more deluged by a flood of new data from current and proposed sky surveys than any other science. In this age of multi-Terabyte scientific archives, scientists need to share the common lessons from different disciplines to make the most of the opportunities available, and to avoid being overwhelmed by the data avalanche.

The Sloan Digital Sky Survey (SDSS) is a multi-institution project to map about half of the northern sky in five wavelength bands from ultraviolet to infrared (see http://www.sdss.org/). When completed (2005), the survey is expected to image over 200 million objects and collect spectra (redshifts) for the brightest 1 million galaxies among these. The SDSS will revolutionize astronomy in a number of ways, but most significantly it will dwarf current astronomical databases in terms of its size and scope. The raw data is expected to exceed 40 Terabytes (TB) and the resulting archive available for scientific research (object catalog) will be 2-3 TB in size. The survey's information content will be larger than the entire text contained in the US Library of Congress.

A detailed description of the SDSS project, including its technological achievements and its goals, appeared in a previous issue [1]. Here we concentrate on the SDSS Science Archive, which will be the final repository of the calibrated data produced by the survey. The Science Archive will be two databases resulting from the survey – a *Target* database holding the catalogs used for spectroscopic targeting, and the *Best* database holding the catalogs generated by the latest (and best) software pipeline. The raw, uncalibrated pixel data (~ 40TB) will be stored in the *Operational* Database (OpDB). The data warehouse layers for both the OpDB and the Science Archive are commercial Database Management Systems (DBMSs).

**Comment [k1]:** No SX At all? Maybe we just shouldn't mention an abbreviation, just Science Archive.

Our initial choice of DBMS was a commercial object-oriented database system (OODBMS, hereafter OODB for short). At the time this decision was made (c.1995), the leading OODBs offered significant advantages for the data model and application that we anticipated for the SDSS Science Archive. Object databases have a larger set of available data-types and the ability to use object associations to traverse references (links) between objects instead of expensive table joins in relational models. These considerations directed our choice of a commercial OODB as the data repository.

The particular OODB product that we chose was selected based primarily on its superior performance, transparent and configurable data organization, and binary compatibility across a range of platforms. In spite of these anticipated advantages, as the size of our data grew, we began to have problems with the OODB in terms of its performance, support, manageability and missing features. These problems and the lack of support from the vendor grew to the point where we could no longer meet the demands of our user community, and we decided to migrate our data to a relational DBMS (hereafter RDB for short).

This conversion from one DBMS to another was a major and traumatic decision for the project, and a lot of work for us. The process is has taken several years and is still not complete – there are still vestiges of the OODB in the project. This article describes why and how we made the conversion.

We begin with an overview of the SDSS Science Archive, followed by a detailed description of our initial implementation of the SDSS Science Archive on UNIX and Linux-based OODB platforms. Then we list the various problems that we encountered with the DBMS that made the SDSS Science Archive virtually unusable as a data-mining tool for the SDSS collaboration and made our administration tasks impossible. This prompted us to look for an alternative. Section 3 is devoted to the object-oriented implementation of the Science Archive. Section 4 deals with our decision to migrate the Science Archive to a relational database and some of the pros and cons of that decision. That section compares the features of the two approaches. Finally, Section 5 compares the performance of the two implementations and includes the results of performance benchmarks.

> **Comment [k2]:** Hmm. That's a little strong. We did have influence on them to make v6.0 faster. What about my rephrasing?

# 2 Science Archive Overview

The SDSS data is collected at Apache Point Observatory. This raw data is first calibrated and fed to a pipeline code at Fermi National Laboratory that performs several operations on it before it is stored in the OpDB (OA in Figure 1). After further processing, 'survey quality' data is exported to the Master Science Archive (MSA). The total data is expected to be a few TB in size. It is accessible to the entire astronomical community through specialized tools, and to the public at large through the Internet. The Science Archive is replicated at SDSS member institutions, with the master archive residing at FermiLab.

> **Comment [jg3]:** Moved to present tense, fixed figure number, and connected paragraph to figure.

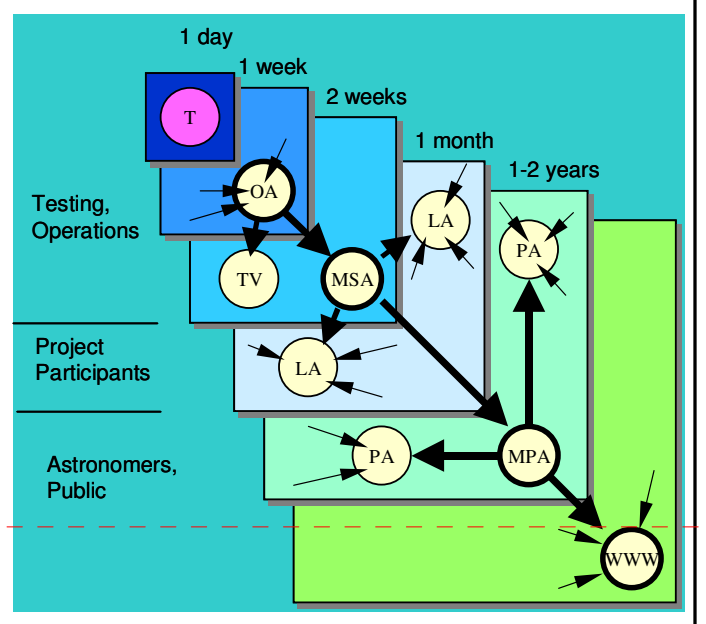

**Figure 1**. SDSS Data Flow. The data is first stored in the operational archive (OA), after which it is calibrated and stored in the master Science Archive (MSA) and local archives (LA's).

## 2.1 SDSS Data Products

The Science Archive contains several data products, including photometric and spectral catalogs, a redshift catalog, images, spectra, links to other catalogs, and documentation. Its data model is optimized for data mining access to the catalogs. Other data products are accessible indirectly. The complete list of data products available through the Science Archive is given in Table 1. The raw data is saved in a tape vault at FermiLab. See Figure 1 for the conceptual SDSS data flow

***Table 1:** Science Archive Data Products*

| *Data Product* | *Size* | *Comments* |
|---|---|---|
| Object catalog | 400 GB | parameters of $>10^8$ objects |
| Redshift Catalog | 1 GB | parameters of $>10^6$ objects |
| Atlas Images | 1.5 TB | 5 color cutouts of $>10^8$ objects |
| Spectra | 60 GB | in a one-dimensional form |
| Derived Catalogs | 20 GB | clusters, QSO absorption lines |
| 4x4 Pixel Sky Map | 60 GB | heavily compressed |

### 2.2 SDSS User Community

The Science Archive needs to effectively handle three types of users:

- **Power Users** are very knowledgeable users within the astronomical community, with lots of resources. Their research is centered on the archive data. Their usage is likely to be moderate numbers of very complex queries in addition to many statistical queries with large output sizes.
- **General Astronomy Public** is a larger community that in aggregate will make frequent, but casual lookup of objects/regions. The archives will help their research, but will probably not be central to it. There will be a large number of small queries and many cross-identification requests.
- **General Public** will essentially be browsing a 'Virtual Telescope' and using the data in education (the education projects of the SkyServer have had over a million web hits in the last year – and these projects created much of the data-access traffic as students used the site to do the 150 hours of online training .) This public access has surprisingly large appeal and makes the SkyServer among the most popular websites at FermiLab (where it is hosted).

## 3 The OODB Implementation of the Science Archive

We first implemented the Science Archive on an OODB. This section describes that implementation and the tools we built to support it. It also outlines some of the challenges we faced once the implementation was complete.

### 3.1 User Access to the Science Archive – the SDSS Query Tool

Fig. 2 shows the primary functional units of the Science Archive from a user's perspective. The astronomical community can access the full data and query facilities of the Science Archive via a user account (username and password). Users start up the Science Archive GUI – the SDSS Query Tool ***sdssQT*** – which is a portable Tcl/Tk client application that communicates with the Science Archive Query Agent over a TCP/IP connection.

The sdssQT allows users to open multiple sessions with different Science Archive servers simultaneously and submit multiple queries to each server in parallel. Queries submitted via the sdssQT can direct their output either back to the GUI (default), or they can send the output directly to a file or an analysis tool via a different socket connection. Binary output capability is also available to allow compact output formats.

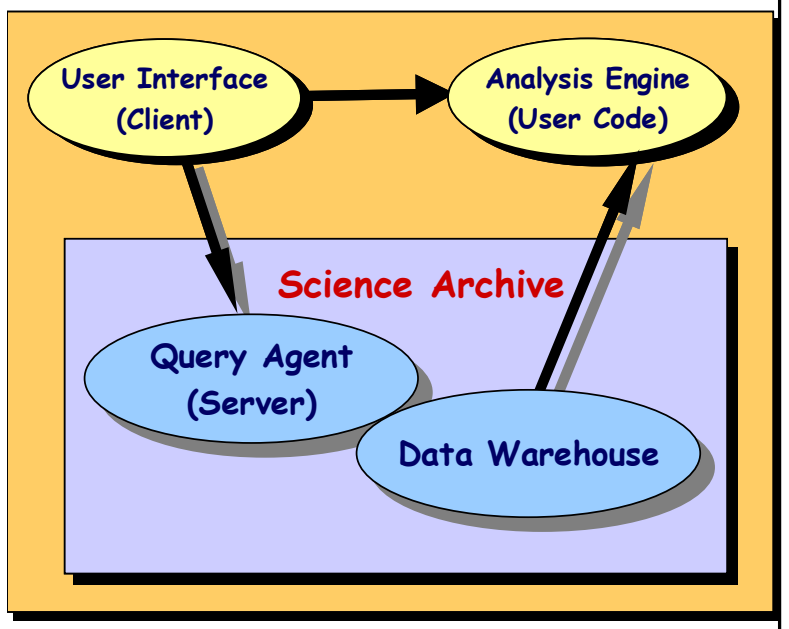

**Figure 2** - Client/Server architecture of the OODB Science Archive. Users can submit queries to the Science Archive Query Agent (server) with a GUI client (sdssQT) and direct output from queries to preexisting analysis engines.

**Comment [k4]:** Through?

**Comment [jg5]:** I thought QT was now deprecated and replaced by SQL QA and by the Emax macros and by the web intereface.

### 3.2 Query Processing in the OODB

Queries presented to the OODB are processed by the Science Archive Query Agent, which is an intelligent query server [3, 4] that a) manages user sessions, b) parses, optimizes, analyzes and executes user queries, c) extracts individual object attributes as requested by the user; and d) routes query output to the target specified by the user. The query agent uses a number of modules to perform its various tasks. The agent architecture and the modules are described in detail below. In essence, we built our own database query optimizer in order to access the OODB database.

### 3.3 Query Analysis

The Science Archive Query Agent first analyzes each query in order to provide a projected "query cost-estimate". The query cost is specified in terms of the subset of the database that must be searched and a rough guess of the time that will be required to complete the search. The user can decide, based on the scope and cost of the query, whether the query is worth running or not.

The projected query cost is computed by first building a **query tree**, and then intersecting the query tree with a pre-constructed **multi-dimensional spatial index – the Hierarchical Triangular Mesh** [5,6]. The intersection yields the scope of the query in terms of the number of *databases* and *containers* that will be touched by the query. An entire (generally distributed) database is referred to as a *federation* in the OODB parlance, and each federation consists of many individual database files. Each database is further subdivided hierarchically into containers, pages and slots. The query tree, along with the scope obtained from the intersection, yields a **query execution tree**, which is essentially the user's parsed query in tree form with scope information included in it. The query execution tree is executed by the query engine to return the objects that are selected by the query.

### 3.4 The Science Archive Query Language

The Science Archive query language, dubbed SXQL (for SDSS Extended Query Language), is a simple SQL-like language that implements the basic subset of clauses and functions necessary to formulate queries for an astronomy object database. There is no attempt to be OQL (Object Query Language) compliant, although we have borrowed some concepts from OQL. SXQL recognizes the SELECT-FROM-WHERE clause of standard SQL, including nested SELECT statements. It further allows specification of *association* links in the SELECT, FROM and WHERE sub-clauses. Associations are links to other objects, and can be either to-one (unary) or to-many (*n*-ary) links. SXQL also recognizes a *proximity query* syntax, which allows the user to search for all objects that are close to a given object in space. Finally, SXQL contains a number of astronomy-specific macros and has full support for mathematical functions.

### 3.5 Tag Objects

We defined a class of lightweight objects to speed query searches. The **tag** class encapsulates the data attributes that are indexed and/or requested most often. Thus, the spatial coordinates and the fluxes are included, as are the status and photometry flags. The tag class is ten times smaller than the whole class and so exhaustive searches run ten times faster. The tag objects were designed to dramatically improve the speed of database searches in the following ways:

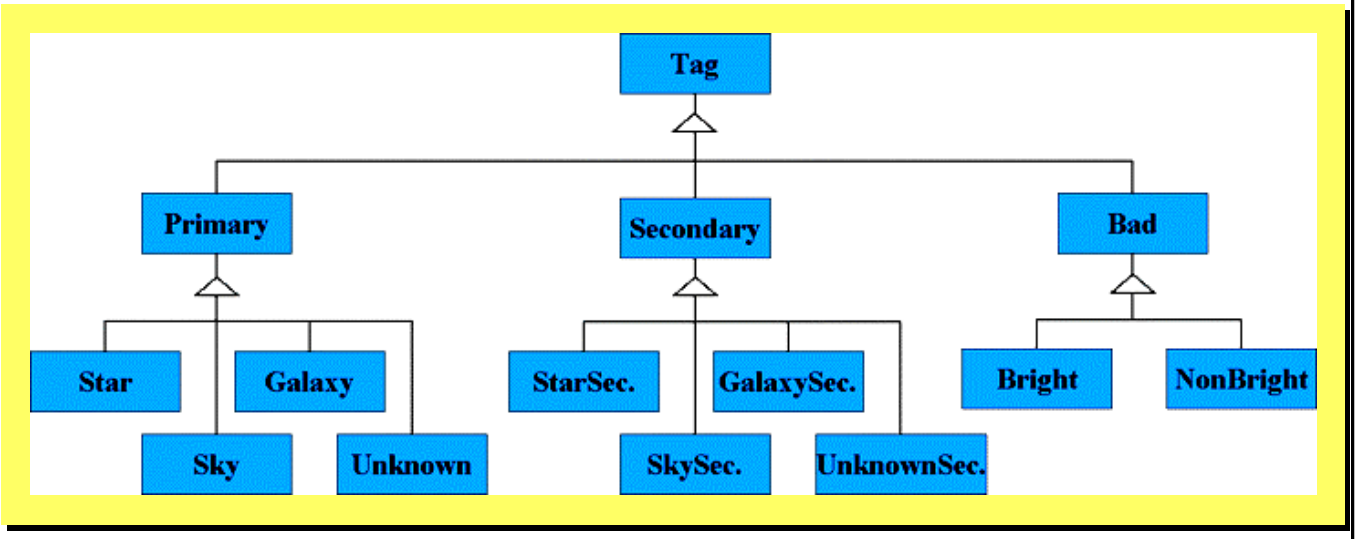

**Figure 3** - The tag class hierarchy narrows searches to the chosen subclass.

**Indexed Lookup:** Encapsulating the most popular data attributes so that indexed lookup can speed up the majority of searches. We develop multi-dimensional spatial and flux indexes [5,6] specifically for this purpose.

**Caching:** The small size of tag objects ensures that many more are loaded into the cache, thereby speeding up the I/O.

**Specialization:** The base tag class, **Tag**, is further specialized into several subclasses, as shown in Fig. 3. Each tag subclass contains exactly the same data members, but the class hierarchy is a powerful way to narrow searches on tag objects. This is because all the tags are stored in separate OODB containers in each database, and different subclasses of tags are stored in different containers. Hence if a search is initiated on Galaxy tag objects, only the Galaxy container in each database file will be searched.

### 3.6 Science Archive Query Agent

The autonomous Science Archive Query Agent – the ***Server*** – processes the queries once a query plan has been chosen. It has the following features:

- it authenticates the user and opens a user session with the GUI client over a socket;
- it maintains multiple concurrent user sessions;
- it allows multiple concurrent queries per user session;
- it routes query results to an Analysis Engine or back to the GUI.

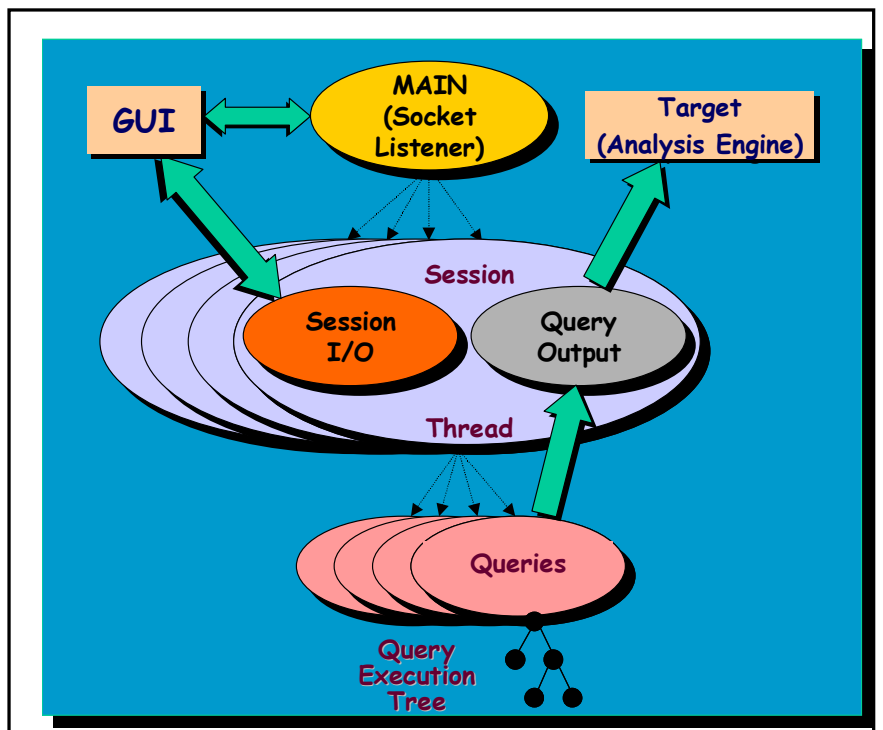

**Figure 4** - The Query Agent thread structure. Each user session is in a separate thread, each query is also handled by a separate thread.

The server software is fully multi-threaded [3]. It incorporates multi-threading at several levels (see Fig. 4). At the top level, the server has a main thread that runs forever and listens on a socket to accept new user connections. Each new user session spawns two threads - one each for input and output from/to the GUI. Searches on remote partitions in a distributed federation are executed in parallel remotely by the multi-threaded remote slave servers (***Slave***). Fig. 3 illustrates the distributed operation of the Query Agent.

The Query Engine executes each query, also in multi-threaded mode, and returns a **bag** of pointers to the selected objects as output. A bag is a generalization of a set in that duplicate members are allowed. The Extractor module is then used to extract selected members from each object in the bag.

### 3.7 Query Engine

The Engine library module implements the query engine, which executes the SXQL query and returns a bag of matching objects. The input to the Engine is the Query Execution Tree (QET). Each node of QET is either a *query* or a *set operation*. The currently supported operations are Union, Intersection, Difference and Distinct. The latter converts a bag to a set. The union is $n$-ary, intersection and difference are binary, and the distinct operation is unary. We use the OODB's predicate query facility to perform the Science Archive query primitives, although we could not use the OODB's predicate-matching and had to develop our own predicate facility, as described below. Each node of the QET is executed in a separate thread, and an **ASAP data push strategy** (data is pushed up the tree using stacks as soon as it becomes available from child nodes) ensures rapid response even for long queries.

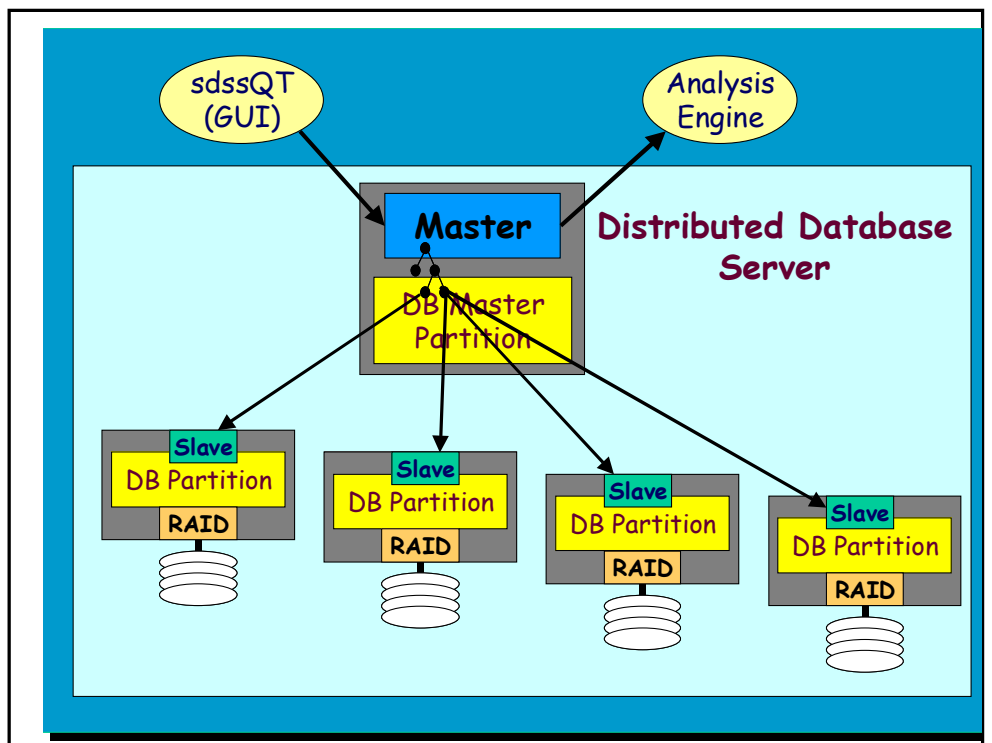

**Figure 5** - Parallel and distributed query execution of Science Archive Query Agent. Searches on database partitions are executed in parallel by slave servers.

### 3.8 <u>Query Execution Tree</u>

An example of a Query Execution Tree is shown in Fig. 6. The nodes in the QET are the Science Archive query primitives and set operations. The QET nodes currently defined are described below. The query nodes map onto the OODB query primitives. Each node returns a bag of object pointers to its parent. All nodes except the leaves have a bag as an input as well as output. The leaf nodes only have an output bag, since they operate directly on a database scope. The different types of query nodes (query primitives) are described below.

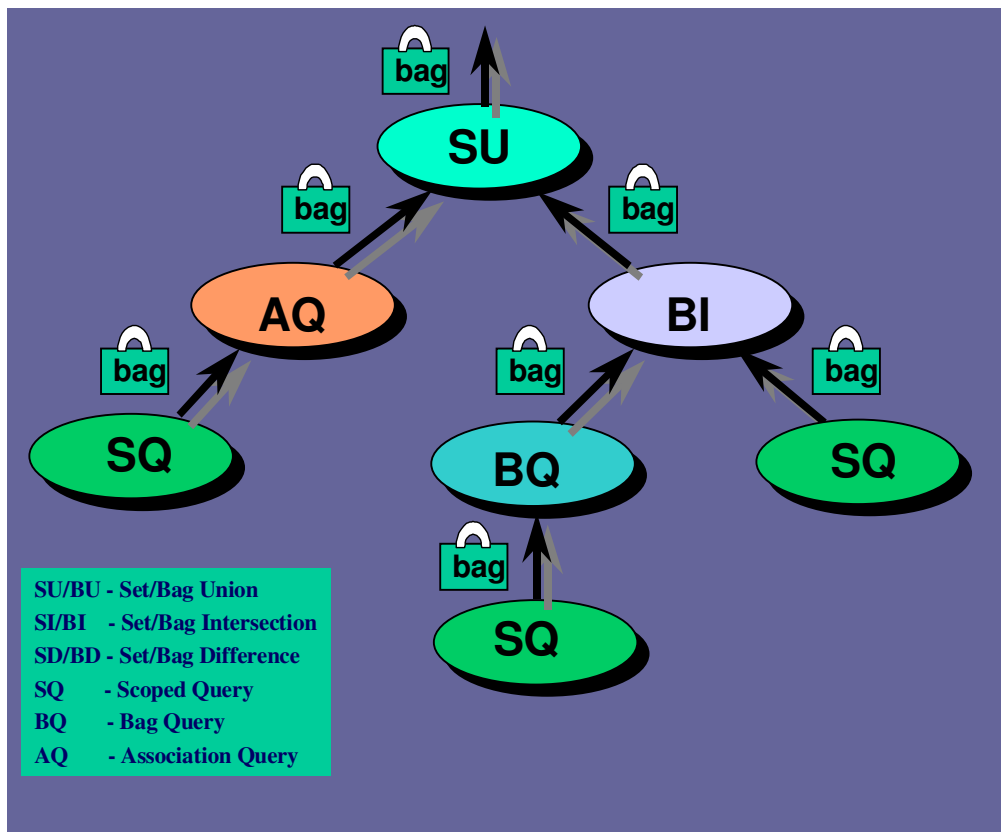

**Figure 6** - Query Execution Tree Example.

**Scoped Query** – This uses the OODB's scoped query functionality. A subset of the federation (scope) is specified for the search in the form of a node-list. The predicate is applied only to the selected type of objects. The scoped query is the only type of query primitive that creates a bag of objects from scratch; hence it is a leaf node in the QET. All other types of queries operate on at least one bag, and they form the interior nodes of the QET.

**Bag Query** – The Bag Query applies the given predicate to each object in the bag that it obtains from its single child node.

**Association Query** – An Association Query allows the selection all objects linked through an association-link to each object in the bag obtained from its single child node. The given predicate is then applied to the selected objects.

**Proximity Query** – This is a search for all objects that are nearby in space to a given object. Such a query is very useful and common in astronomy.

### 3.9 <u>Overall OODB Science Archive Architecture.</u>

**Several** other modules that play important roles in the Science Archive machinery. Figure 7, cartoons the Science Archive software architecture, including the user interface or client layer, the query support or server layer, and the data warehouse layer. The following sections give a synopsis of these other components

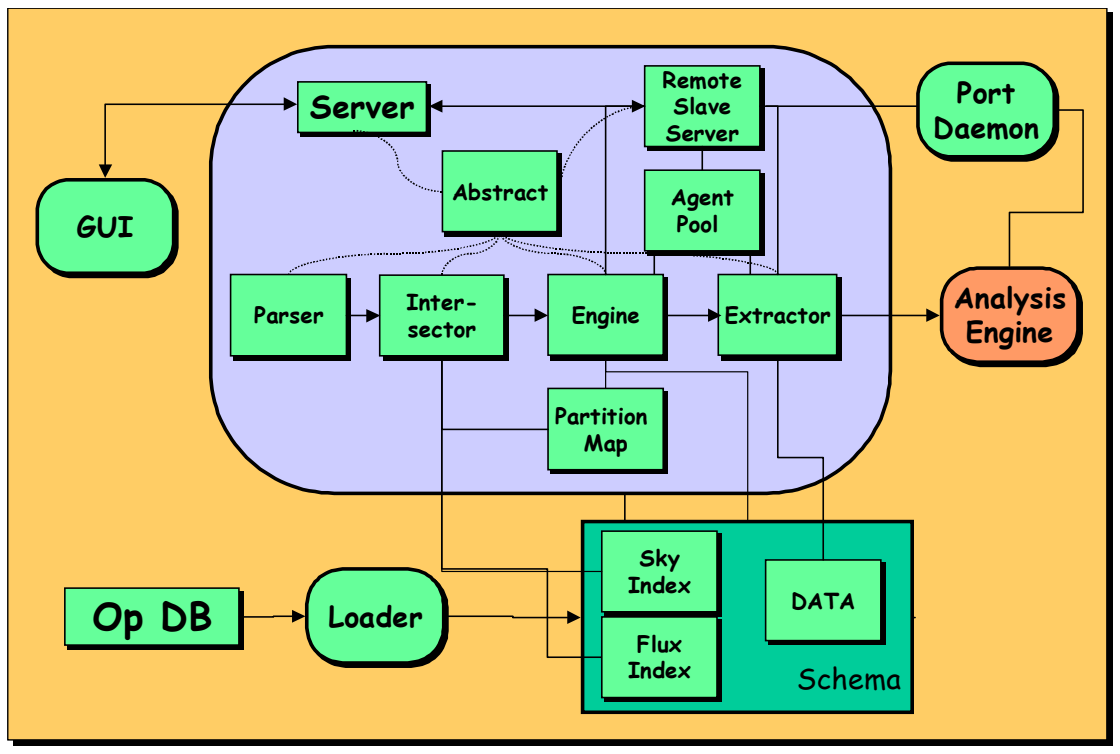

**Figure 7** – Overall OODB Science Archive architecture.

#### 3.9.1 Parser, Intersector and Mapper

The ***Parser*** module is responsible for parsing the SXQL query and converting it to a query tree, which is then passed to the ***Intersector***. The Intersector intersects the query tree with the ***Spatial*** and ***Flux Indexes*** [5,6]. The ***Partition Map*** tells the query agent how the data is distributed in the federation by identifying partitions that are on local and remote partitions. This enables data on different partitions to be searched in parallel.

### 3.9.2 Abstract

The Abstract library module provides a run-time abstraction of the Science Archive data model. It allows manipulation of DB objects without knowledge of schema, retrieval of data values, and invocation of methods (if applicable). In effect, it behaves as a runtime metadata server or type manager It provides functions to:

- specify an object's class using a name or alias
- translate a name or alias to an actual C++ class
- identify a particular member of a persistent class and retrieve the following information about it:
  - the kind of member (basic data, array, function, association link)
  - type of value returned by member (int, float, string …)
  - size of array or object member
  - input and output formats
- return a pointer to the actual object in memory.

The Abstract can be used by any application/module that accesses the data model. It is used by the Parser, Intersector, Query Engine and Extractor modules.

### 3.9.3 Extractor

The Extractor library module provides the functionality needed to extract individual or groups of attributes (members) of a given object (OID). It therefore executes the SELECT part of the SXQL statement. The members that it can extract include data, function and association-link members. The Extractor retrieves data values as per information provided by the Abstract. It has the required mechanisms for invocation of object methods.

### 3.9.4 Port Daemon

The Port Daemon ensures that the GUI and Analysis Engine communicate with the Server on the correct port and also performs process-level authentication. This is also necessary for security. Firewalls are opened only for the port daemon's own port and the predetermined port(s) configured for the Server process.

### 3.9.5 Data Loader

The Data Loader and is the application that loads new data into the OODB federation from the SDSS data pipeline. The SDSS data flow requires only infrequent updates at well-defined intervals – when new data is collected and calibrated. These bulk additions are done offline by the Data Loader. The Server opens the federation in read-only mode and the OODB takes care of concurrency issues. The incoming data format is hard-coded into the Loader module to prevent data mismatches. The latest version of the data loader also incorporates the ability to store different calibrations (versions) of the same data that need to be accessible through the archive.

## 3.10 OODBMS Evaluation

### 3.10.1 Advantages

Early in the planning of the Science Archive, we decided that an object-oriented data repository would offer significant advantages over a relational or table-based system. In addition to the conceptual attractiveness of a DBMS that stored the data in the same way that astronomers thought about it, there were several other considerations that made the choice of an OODB in general a beneficial one:

**Object associations** allow designers to **link objects** to each other – the OODB allows one-to-one, one-to-many and many-to-many links between objects. These are called **associations**.

**Transparent and configurable organization** of data on disk – the OODB has a multi-layered hierarchical data organization with the data **federation** at the top level, followed by **database**, **container**, **page**, and **slot**. A federation is a distributed data repository in general, and can contain thousands of databases. Each level in the hierarchy can contain up to 65k objects of the next level. Thus each database can contain up to 65k containers, and so on. More importantly. **Each database is a disk file** and the placement of objects into containers is configurable by the database administrator.

**Binary data compatibility across platforms** – the OODB database and federation files can be moved across heterogeneous platforms without loss of compatibility. This greatly facilitates the implementation and maintenance of distributed archives.

**Object methods** – the OODB was expected to make available the capability to define class methods that could be stored in the database along with the object data. This would essentially be stored procedures, but on a per-class basis. This feature is, however, still not available in the latest version of the OODB, although we are able to provide macros and object functions via our query language.

**Query tools** – the OODB community was developing OQL and promising non-procedural access to in a standard query language.

**Development tools:** The OODB products had good tools that dovetailed with our chosen development environment based on C++.

**Administrative tools**: The OODB products had a complete set of utilities for data definition, data security, and data administration.

**Data replication tools** – this is the ability to make copies of the data archive and automatically propagate updates to archive mirror sites.

The first five features facilitate data-mining performance and make it possible to run efficient queries for "needle-in-a-haystack" type of searches.

### 3.10.2 OODB Shortcomings

Even during the survey commissioning phase, it became clear that the SDSS would be using the OODB in a regime where it had never been tested before. The data mining features and query performance that the SDSS user community demanded from the OODB were not required in the other application domains that the OODB was used in. The problems became progressively worse as our user community got more sophisticated and familiar with the query language. The features and functionality that we found lacking include the following.

#### *3.10.2.1 Inadequate query language*

The query language supplied with the product falls far short of what was promised 15 years ago. The current version of the OODB supports only Boolean user-defined operators, i.e., the only value returned by the operator is either *true* or *false*. These are too restrictive. It would be nice to have more versatile operators that can perform arbitrary complex tasks and return values of arbitrary type. Operators for manipulating individual bits of data members are not supported in query predicates. This functionality would be very useful to us given our extensive use of bit-lists for our indexing. The ability to directly invoke member functions of persistent classes would allow us to store analysis and processing code in the archive itself instead of running these methods manually after retrieving the objects. In general, the query language was inadequate, so we had to build our own atop the OODB with the incumbent implementation and execution costs.

*3.10.2.2 Bugs*

There were several serious bugs that we found over the three years of using the OODB that are still unresolved at the time of this writing – we had to devise workarounds for each of them at considerable development cost to us. These are described below. The most critical bug that we encountered was incorrect array addressing of single-precision (float32) floating-point array members in query predicates. For example, if *f* is a floating-point array and the user asks for *f*[*n*], i.e. the *n*-th element of *f*, the OODB returns *f*[2*n*], i.e. the array offset is off by a factor of 2! This happens only with float32 arrays, not with float64 arrays, which suggested that the problem most likely was that the OODB assumed every floating-point array to be float64.

Fatal errors generated during a query automatically caused the C++ application to abort with a core dump. Defining an error-handler did not help because the error-handler could not avoid the abort – it could only report that a fatal error had occurred. This was very inconvenient indeed because a fatal error caused our entire query agent to crash, throwing off all users logged in at the time even if the error was generated by one particular user. For example, we encountered a situation where a division in the query predicate caused a fatal error (floating exception) if the value of the object attribute that is in the denominator happens to be 0 for one of the objects encountered in the search. Although it is reasonable to expect the query to be aborted by such an error, it is ridiculous to have to abort the query agent because of it. There appears to be no way to isolate the error to the thread or query that generated it and continue with the remaining threads. We have not received a satisfactory resolution of this problem from the OODB vendor to date.

The missing features and critical bugs listed above made the OODB's predicate query feature unusable for our application. The lack of reliable predicate-query functionality obviously crippled our ability to make effective queries against the Science Archive, and was a serious blow to our progress with the Science Archive development. As a result, we had to write our own predicate functionality, which required several person-months of precious development effort. Our home-cooked predicate query had served us very well until that point and even allowed us to implement features not available in the OODB (such as inclusion of object methods in query predicates), but certainly placed an additional burden on the support and maintenance of the software.

*3.10.2.3 Performance Woes*

To serve our customers, we needed to be able to access data at more than 50MBps – that would be an hour for a query that examines the entire photo catalog. No matter how much hardware we threw at it, the OODB could not go much more than 0.5 MBps – 100 times too slow for our needs. We eventually traced this to the OODB's use of NFS and the lack of high-speed sequential access to the data at the file system level. There was now way around this without reworking the OODB – something the vendor refused to consider.

Another serious problem was that if indexes were defined on several quantities in our tag object class, the order in which the terms involving these quantities appeared in the query predicate <u>must be the same as the order in which the indexes were created</u>; otherwise none of the indices was applied! For example, the predicate,

```
(i < 11 && r < 13 && z < 9),
```

where *i,r* and *z* are members of the tag object, will generate a fast (indexed) search only if the index on *i* was created first, the index on *r* was created second, and the index on *z* was created last.

*3.10.2.4 Administrative Problems*

Inevitably, we had to change the database schema, adding attributes or classes, creating associations, and making other schema changes. As often as not, this required a complete database reload (and indeed the online schema change often corrupted the current system). One thing the OODB community has yet to address is a coherent strategy for evolving the database schema when the application programs have such a tight coupling to the data and when the data has object pointers that may be "spit" or "merged" by a schema change.

In addition, the OODB lacked the administration tools typical of more mature database management systems. In fairness, other projects, notably BaBar succeeded in building atop an OODB, but they did this by writing over a million lines of code above the interface, and by using the OODB as a persistent object store (checkin-checkout data) rather than as a data warehouse with ad-hoc access.

# 4 Evaluation of and **Migration to RDBMS**

In light of the issues discussed above, it became increasingly difficult to justify the use of the OODB in spite of the conceptual attractiveness it offered and the fact that we had invested so much time and effort in building a distributed data-mining engine on top of it. Apart from the issues relating to the particular product that we had deployed the Science Archive on, object persistence itself was proving to be too expensive for us because of the poor query language support, the relative inflexibility of the data model to frequent changes, the total lack of query optimization, and last but perhaps most debilitating for the purposes of data mining – inadequate I/O efficiency and support for data striping.

Happily for us, in the last ten years the landscape has shifted. Relational database management systems (RDBs for short) have improved in the following ways:

- RDBs have matured as the default database technology embraced by the computer industry. This has meant wide tools support, dramatically improved quality and much improved functionality in areas like language integration and database extension mechanisms that provide some of the benefits promised by OODBs.
- The fierce competition among the major RDB vendors (e.g. Oracle, IBM, and Microsoft) has resulted in excellent I/O optimization. Ten years ago, when we decided to adopt an OODB for the Science Archive, object-oriented database technology *was* in fact superior in performance to the existing relational technology at the time. However, in the last few years, the leading object database vendors have not kept pace with the order-of-magnitude increase in raw I/O performance achievable with RAID disk systems.
- The advent of web services, SOAP and XML enables objects-on-the-fly, i.e., retrieval of data in the form of objects irrespective of how the raw data is stored on disk, providing dynamic object interfaces to relational databases and making object persistence unnecessary.

These developments prompted us to consider migrating the SDSS Science Archive to a commercial RDB to gain significantly better performance, database administration and optimization.

## 4.1 <u>The SDSS SkyServer</u>

Our original motivation in deploying a version of our data in MS-SQL was to provide an easier to use, web-based option – called the **SkyServer –** for the casual user of the soon-to-be-available first public distribution of the SDSS (the Early Data Release or EDR), which was officially released in June 2001. The choice of Microsoft's products was dictated by the skills available – Microsoft was willing to support the project, and one of us (Gray) was intimately familiar with the product. In the absence of that we would probably have used Apache/DB2 on Solaris.

In order to load the EDR data into the SkyServer, we had to modify our data loader module so that it exported the data in CSV (ASCII comma-separated values) format for loading it into the relational data tables. The original version of the data loader was designed to feed the data from the data pipeline directly into the OODB using the OODB proprietary Data Definition Language (DDL). The addition of the CSV export feature to the data loader proved not to be too difficult or time-consuming, and we were able to have an MS-SQL EDR database up and running in a matter of weeks. This CSV interface provides a blood-brain barrier between to the two subsystems and has been invaluable in allowing each component to evolve independently.

The SkyServer was designed to be a user-friendly version of the EDR database. It included help and documentation for novice users, and was pitched to the lay public, students, and amateur astronomers rather than professionals. Although the interface included a SQL query submission facility, it was geared more towards the casual user interested in browsing the SDSS data rather than running intensive queries for data mining. We did not expect the SQL query page to be heavily used.

We were wrong. The SkyServer stole the show during the first 6 months of the EDR. Users quickly realized that this was an easy way to get data out of the EDR. Some groups used the interface to crawl the entire site and get a complete extract. Others wrapped the query interface in a collection of Emacs macros and would routinely get personal extracts of the data. Still others got private copies (we distributed about 40 copies of the full EDR and several hundred copies of the 1% subset called the *Personal SkyServer*.) Our highly positive experience with the EDR SkyServer, in terms of its popularity even with professional astronomers due to its versatility, reliability (99.9% uptime) and performance convinced us that it was a viable alternative as an advanced data mining facility. This was in contrast to the persistent and critical problems that we were experiencing with the OODB version over the same period, particularly in terms of stability, reliability and performance. Indeed, with major data model changes just around the corner, the SkyServer SQLserver system emerged as the far more attractive option (see the following section). But this was not a decision to be taken lightly. Consequently, we undertook a comparative evaluation of the OODB and Microsoft SQLserver with performance benchmarks to formally ascertain whether SQL would meet our demanding needs (section 8).

### 4.2 Moving to the Relational Data Model

The main hurdle we had to overcome was translating our object data model into a relational data model. The relational model is "flat", there are no sub-classes. The relational model does not support "pointer" associations; all associations must be via key values. The hierarchical object data architecture had to be converted to relational tables. Beyond the obvious loss of conceptual ease that the object data model gave us, this transformation actually did not prove to be all that difficult. We were able to use views to capture the subclass concept. For example, the Galaxy view is the subset of photo objects that are classified as galaxies and are primary objects. Associations were modeled with foreign keys. Data import was not difficult, partly because we had already written a data loader for our OODB that read the astronomical data in its original FITS format and stuffed it into the OODB using its DDL (data definition language). We were able to simply modify this data loader to also export the data in a format accepted by MS-SQL – CSV (comma-separated value) input files for importing into the relational tables.

There were a few benefits of the object data model that we were sorry to part with when we moved to SQL:

- In OQL, links between objects that can be specified with a straight-forward dot syntax, e.g. "`SELECT obj.field` …" instead of the much less intuitive `JOIN` syntax of relational queries.
- SQL does not support array (vector) fields. This is especially troublesome since most of our fields are measured in 5 wavelength bands. It was really convenient to be able to have an array[5] of each quantity instead of 5 columns in the table for each quantity measured in all bands using the terminology f_u, f_g, f_r, f_i, f_z for the ugriz bands of the survey.
- Our simplified SQL was less imposing and confusing to the user community than standard SQL.

However, the loss of these OODB advantages was not a high price to pay for the considerable advantages of MSQL that we acquired. These are enumerated in the following section.

## 4.3 MS-SQL Advantages

The following is a list of the main benefits that we gained by migrating to MS-SQL.

### 4.3.1 Performance and Stability

As the performance benchmarks described in the following section demonstrate, MS-SQL query performance for a representative sample of astronomical queries is an order of magnitude better than the best performance that we had been able to squeeze out of the OODB after all the optimization and parallelization that we designed and implemented in our query engine. Loading performance is also much faster with MS-SQL, and we can load our current data in several hours with MS-SQL rather than several days as required for the OODB. Subsequently, we have achieved another tenfold speedup in most of these areas by better design and more modern hardware. We are nearing data rates of 1 GBps and load rates of 50MBps.

### 4.3.2 Query Language

In contrast with the limited SQL functionality that we were able to implement in SXQL, MS-SQL offers full SQL92 compatibility and the added features of Transact SQL (T-SQL). The MS-SQL query language contains many convenient features that our user community has been asking for but we have not been able to implement in SXQL due to our limited software development budget. The three most useful features are GROUP BY, ORDER BY, and COUNT(*). The lack of them in SXQL already hampered us in our development of a suite of test queries for our performance benchmarks. We were unable to translate several of a set of test queries – originally developed for a different purpose – to SXQL. SDSS users would love to be able to sort, group and bin query results, thereby saving a fairly time-consuming data analysis step.

### 4.3.3 Advanced query optimization

Some of MS-SQL's greatest strengths are its ease of optimization and self-optimizing capabilities. In a feature article in the February 2002 issue of PC Magazine, MS-SQL scored higher in this aspect than all of its major competitors – but this is a core strength of all the major SQL products. The graphical query plan viewable before submitting a MS-SQL query gives details on exactly which steps of the query take the largest fraction of the execution time, giving the user all the information necessary, in most cases, to improve the query performance. Such a capability was totally lacking in the OODB.

### 4.3.4 DB Administration Features

The availability of features like triggers, stored procedures, dynamic indices, online reorganization, data striping, job scheduling, self-tuning, and many other utilities in MS-SQL provide a level of convenience and ease for DB administration that we did not have with the OODB. In addition, many optimization tasks, including index management and data re-organization, can be done live, i.e. while the DB is online. While this is also possible in principal with the OODB, in practice we have found this to be problematic enough as to be virtually unusable.

### 4.3.5 Schema Design and Evolution

MS-SQL provides built-in tools for data model design that facilitates development of complex schema with great ease. With the OODB, we had to use a separate software product (Paradigm Plus) to develop a diagrammatic data model before translating it into schema files that the DBMS could understand (DDL).

The benefits of schema development in MS-SQL were brought into focus for us when we contemplated major data model changes in order to include detailed information about the tiling procedure for the SDSS spectroscopic plates into the OODB. We realized that adding these changes would stretch our existing database to the breaking point.

Schema evolution is another feature that we should have been able to take for granted with a DBMS but have been unable to use in practice with the OODB. Even small schema changes ended up corrupting the existing data in the database, requiring a complete reloading of the data. After several unsuccessful attempts at schema evolution, and lacking adequate technical support from the vendor, we were forced to abandon schema evolution altogether. Instead we kept schema changes to a minimum and always reloaded the entire database when a change was absolutely necessary. Naturally, this was a significant drain on our limited operations resources since it required several days to implement even a minor schema change. We anticipate that this will no longer be a major problem with MS-SQL, and have already done several schema changes successfully with our EDR SkyServer.

#### 4.3.6 XML Compatibility

XML output capability is available in MS-SQL now: simply by specifying FOR XML in the WHERE clause of the query, the user can opt to receive the query results in XML rather than ASCII text. This is an especially important feature for the upcoming Virtual Observatory [7], in which many of the data analysis applications that astronomers need will most likely be available as web services. We would have to build XML compatibility ourselves on top of the OODB.

# 5 Performance Evaluation

Performance was not the only reason for switching to MS-SQL, but it was probably the most important one. With the data volume expected to quadruple over the next 3 years, we were justifiably concerned that all the optimization that we had built into our OODB query engine would not be enough if the DBMS itself did not meet certain IO performance benchmarks. Almost all the queries that had been performing poorly on the OODB were IO-bound.

Before we carried out the performance benchmarks described here, we already knew that our SQL Server database – the SDSS SkyServer – was performing considerably better than the OODB version for certain types of queries. However, we conducted formal benchmarks for the following reasons.

- To satisfy ourselves that MS-SQL was indeed performing much better with indexed queries.
- To ensure that we were giving the OODB a fair shake. We wanted, if possible, to find complex queries that the OODB would perform better with due to its hierarchical data organization.
- To provide a level playing field by running the benchmarks on identical hardware.
- To make every attempt to ensure that the optimization and parallelization features that we had built into our OODB server software would be utilized in executing the test queries.

Therefore, we formulated a set of 25 test queries that range from simple index lookups to queries with complex constraints on non-indexed quantities and multi-way joins. These queries – listed at http://www.sdss.jhu.edu/sx/pubs/ – are a subset of 35 queries used to benchmark the SkyServer [8].

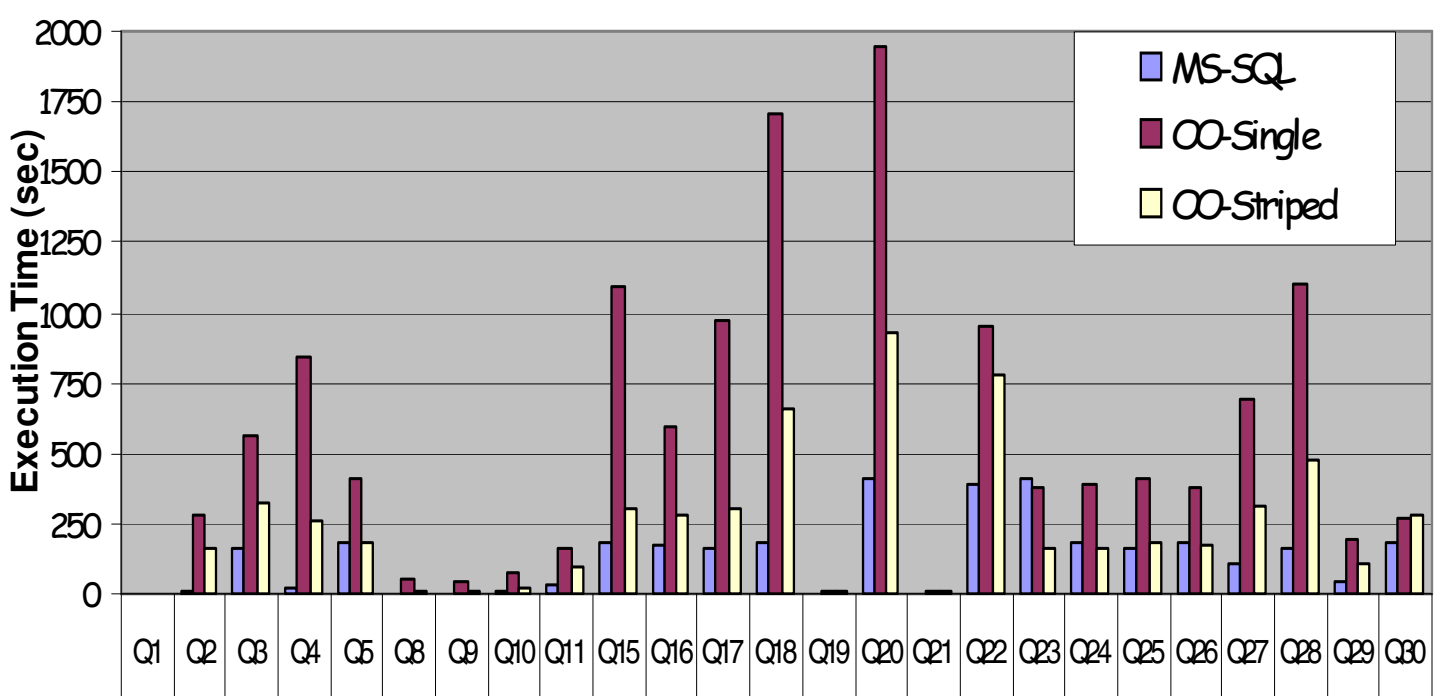

| | Q1 | Q2 | Q3 | Q4 | Q5 | Q8 | Q9 | Q10 | Q11 | Q15 | Q16 | Q17 | Q18 | Q19 | Q20 | Q21 | Q22 | Q23 | Q24 | Q25 | Q26 | Q27 | Q28 | Q29 | Q30 |
|---|---|---|---|---|---|---|---|---|---|---|---|---|---|---|---|---|---|---|---|---|---|---|---|---|---|
| MS-SQL | 0.2 | 13 | 157 | 22 | 184 | 1 | 2 | 9 | 36 | 185 | 169 | 158 | 181 | 3 | 409 | 3 | 390 | 409 | 188 | 157 | 184 | 104 | 163 | 39 | 182 |
| OO-Single | 1 | 286 | 559 | 848 | 409 | 52 | 48 | 73 | 161 | 1090 | 595 | 969 | 1709 | 15 | 1946 | 15 | 950 | 375 | 388 | 416 | 383 | 693 | 1098 | 194 | 271 |
| OO-Striped | 1 | 161 | 327 | 262 | 186 | 12 | 11 | 26 | 97 | 298 | 283 | 301 | 659 | 13 | 933 | 15 | 783 | 160 | 160 | 180 | 175 | 314 | 481 | 104 | 280 |

**Figure 8:** Results of running the 25 test queries listed in the appendix on Microsoft SQL Server 2000 (MS-SQL), on a monolithic OODBS (OO-single), and on a data-striped OODBMS (OO-Striped). The tests ran on identical hardware using the SDSS EDR data (about 80GB). The reported times are in seconds.

## 5.1 Analysis of Performance Benchmark Results

The results of execution time measurements for the test queries are shown in Fig. 8 for the three DB configurations that we tested – MS-SQL, a single-disk OODB (OO-Single), and a 3-disk partitioned OODB (OO-Striped). The performance differences between the three tested systems are summarized below.

On average, MS-SQL is 3-10 times faster than OO-Single, and 1-3 times faster than OO-Striped. This is shown more clearly in Fig. 9, which compares the execution times on the OO-Single system and the times on the OO-Striped system to MS-SQL. The 3-way striping of the data on OO-Striped yields, as expected, up to a three-fold performance improvement due to our multi-threaded query agent design.

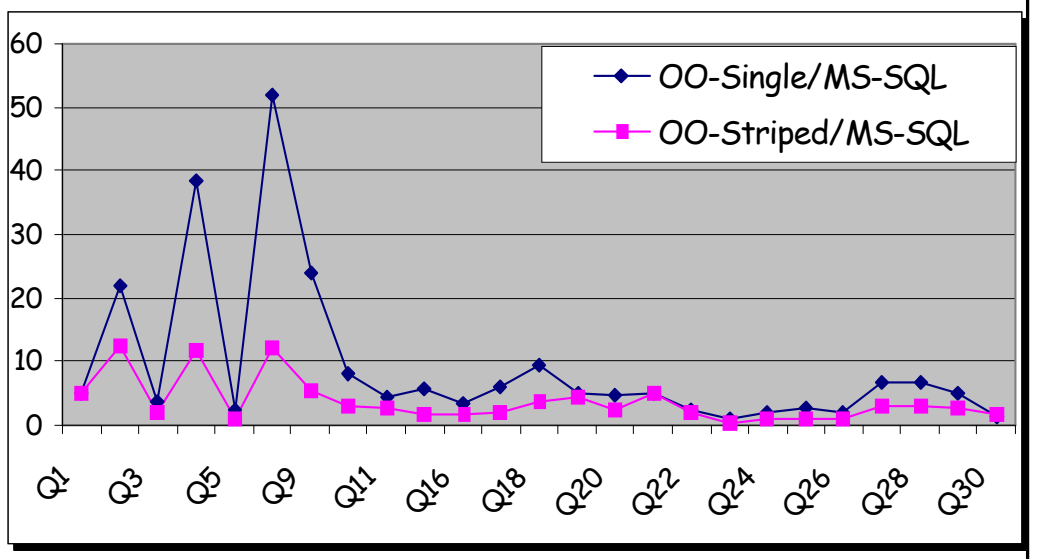

**Figure 9** - Comparison of MS-SQL query execution times with the single-disk OODBMS (OO-Single/MS-SQL) and three-disk OODBMS (OO-Striped/MS-SQL).

On queries that return many objects, the distributed version of the OODB (OO-Striped) starts to bog down on the extraction of the requested data fields from the matching objects. This is especially true when the query is on tag objects, because many more of the smaller tag objects are loaded into the cache at one time, and the main thread has to extract the required data fields from them so that the CPU usage of the main thread approaches 90+% cpu utilization. This limits the speed at which the query is executed. This is illustrated in Fig. 10, which shows the ratio of the times taken to return just an aggregate count of the objects (with COUNT(*)) to the total time taken (to return all objects) for each query. The ratio is lowest, i.e., the count is much faster than the whole query, when the total number of objects is in the millions. This is a limitation of the distributed design of our query agent, which distributes the query search but not the extraction of the object data.

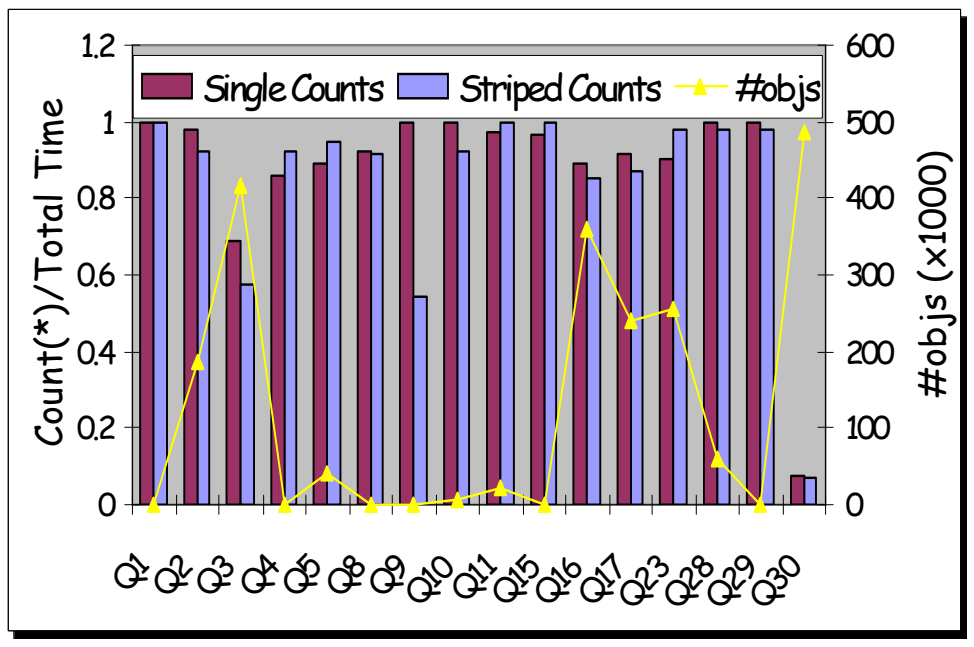

**Figure 10** - Comparison of time saved by requesting aggregate counts as opposed to total time for the query for the OO-Single and OO-Striped versions with respect to the total number of objects returned by each query.

The joins on the MS-SQL version were very fast for all the queries we formulated. In fact, we were hard put to find a query that would take more than a few minutes to execute on the MS-SQL version. The only time when the execution was relatively slow was when there was no constraint on an indexed quantity in the query predicate. An example of this situation is shown in test query Q31, which is a version of Q30 with a much larger limiting value of a non-indexed field in the **photoobj** table. The difference in the execution times is shown in Fig 11 for MS-SQL, OO-Single and OO-Striped. The execution times for Q31 are almost 2 orders of magnitude higher. However, such a query is rare because usually it is possible to narrow the search using one of the indexed fields.

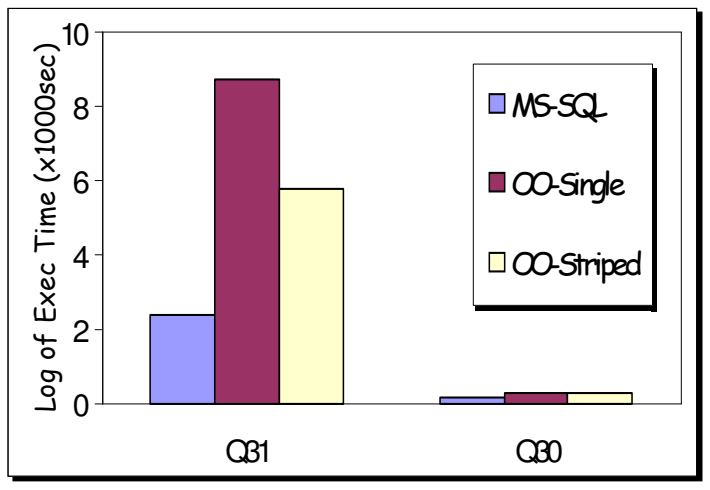

**Figure 11** - Comparison of execution times for two versions of the last test query (Q30). Q31 is a version of Q30 with a much larger limit on a non-indexed quantity.

In several test queries that involve reddening-corrected u,g,r,i,z magnitudes, the MS-SQL version benefits from the fact that for the primary object table (from which Galaxy, Star etc. are derived), the ugriz values are already reddening-corrected. This provides approximately a factor of 2 improvement for the MS-SQL version.

# 6 Summary

After initially adopting an object database and expending considerable time and development effort in converting it into an advanced data mining facility, we were forced to abandon it and migrate to a relational database system instead. The OODB concept was great, but the actual implementation lacked the robustness, the tools, and the scalability that our project demanded. The detailed description of our object-based implementation above bears testament to the concerted attempt that we made to achieve the data mining objectives of the SDSS Science Archive with an OODB. However, in the end we had to acknowledge that relational database technology and web-based protocols have advanced to the point where many of our original reasons for adopting an object database can be met with a relational solution, and the dominant manageability, quality, performance, and usability requirements mandate that we use a mainstream product.

# Appendix

## Test Queries for Benchmark

We used a set of 25 different test queries for the performance benchmarks discussed in Section 5. These are the queries Q1-Q30 listed below (Q31 is a modified version of Q30). The query numbering is not consecutive because this is a subset of 35 test queries that we devised to test the SQL Server's data mining capabilities and query performance in a previous paper. The queries from the original set that could not be converted to SXQL were dropped (e.g., Q7, Q12, Q13), and where possible SQL-specific features were excluded to make the SQL and SXQL versions equivalent. In such cases, the excluded SQL code is shown highlighted below. Both the SQL Server (SQL) and the OODBMS (SXQL) versions are shown side-by-side. The SQL Server versions include a workaround for the SQL Server 2000 optimizer problem that causes the performance to degrade when user defined functions that lookup flags are included in the query predicate (the function is called for each record rather than being called once for the whole query). This is a known bug that will be fixed in the next version, and the workaround is to manually factor the function calls out of the queries (or just use the constants).

1. **Q1**: Find all galaxies without saturated pixels within 1' of a given point (ra=75.327, dec=21.023).

**Q1.SQL**

```
SELECT G.objID, G.ra, G.dec, GN.distance
FROM      Galaxy as G,
   JOIN  dbo.fGetNearbyObjEq(200,-0.5,1) as GN
   ON G.objID = GN.objID
WHERE
   (G.flags & dbo.fPhotoFlags('saturated') = 0
order by distance
```

**Q1.SXQL**

```
SELECT  objID, RA(), DEC()
FROM  Galaxy
WHERE
   PROX( J2000, 200, -0.5, 1.0)
   && (objFlags & OBJECT_SATUR) = 0
```

2. **Q2**: Find all galaxies with blue surface brightness between and 23 and 25 mag per square arcseconds, and -10<super galactic latitude (sgb) <10, and declination less than zero.

**Q2.SQL**

```
SELECT objID
FROM Galaxy
WHERE
 ra BETWEEN 160 AND 180 AND dec < 0
 AND (g+rho) BETWEEN 23 AND 25
```

**Q2.SXQL**

```
SELECT objID
FROM Galaxy
WHERE
  RA() BETWEEN 160 AND 180  && DEC() < 0.0
  && (g+rho) BETWEEN 23 AND 25
```

3. **Q3**: Find all galaxies brighter than magnitude 22, where the local extinction is >0.175.

**Q3.SQL**

```
SELECT objID
FROM Galaxy
WHERE
  r + reddening_r < 22 AND reddening_r > 0.175
```

**Q3.SXQL**

```
SELECT objID
FROM Galaxy
WHERE
  r < 22 AND reddening[2] > 0.175
```

4. **Q4**: Find galaxies with an isophotal surface brightness (SB) > 24 in the red band, with an ellipticity >0.5, and with the major axis of the ellipse having a declination of between 30 and 60 arc seconds.

**Q4.SQL**

```
SELECT ObjID
FROM Galaxy
WHERE
  modelMag_r + rho < 24
  AND  isoA_r BETWEEN 30 AND 60
  AND  (power(q_r,2) + power(u_r,2)) > 0.25
```

**Q4.SXQL**

```
SELECT objID
FROM PhotoPrimary
WHERE
  objType == OBJECT_GALAXY &&
  modelMag[2] + rho < 24
  AND  isoA[2] BETWEEN 30 AND 60
  AND  (q[2]*q[2]) + (u[2]*u[2]) > 0.25
```

5. **Q5**: Find all galaxies with a deVaucouleours profile (r¼ falloff of intensity on disk) and the photometric colors consistent with an elliptical galaxy.

**Q5.SQL**

```
DECLARE @binned    BIGINT;
SET     @binned = dbo.fPhotoFlags('BINNED1') +
                  dbo.fPhotoFlags('BINNED2') +
                  dbo.fPhotoFlags('BINNED4') ;
DECLARE @blended   BIGINT;
SET     @blended = dbo.fPhotoFlags('BLENDED');
DECLARE @noDeBlend BIGINT;
SET     @noDeBlend = dbo.fPhotoFlags('NODEBLEND');
DECLARE @child     BIGINT;
SET     @child =  dbo.fPhotoFlags('CHILD');
DECLARE @edge      BIGINT;
SET     @edge =  dbo.fPhotoFlags('EDGE');
DECLARE @saturated BIGINT;
SET     @saturated =  dbo.fPhotoFlags('SATURATED');
SELECT ObjID
 FROM Galaxy as G   -- count galaxies
 WHERE lDev_r > 1.1 * lExp_r -- red DeVaucouler fit
                    -- likelihood greater than disk fit
   AND  lExp_r > 0  -- exp disk fit in red band > 0
 -- Color cut for an elliptical galaxy courtesy of James
 -- Annis of FermiLab
   AND (G.flags & @binned) > 0
   AND (G.flags & ( @blended + @noDeBlend + @child))
                                           != @blended
  AND (G.flags & (@edge + @saturated)) = 0
  AND G.petroMag_i > 17.5
  AND (G.petroMag_r > 15.5 OR G.petroR50_r > 2)
  AND (G.petroMag_r > 0 AND G.g>0 AND G.r>0 AND G.i>0)
  AND ((G.petroMag_r-G.reddening_r) < 19.2
  AND (G.petroMag_r - G.reddening_r < (13.1 +
   (7/3)*( G.g - G.r ) + 4 *( G.r - G.i )- 4 * 0.18 ))
  AND ( ( G.r - G.i - (G.g - G.r)/4 - 0.18) < 0.2 )
           AND ( ( G.r - G.i - (G.g - G.r)/4 - 0.18)
                     > -0.2 ) )
           OR (  (G.petroMag_r - G.reddening_r < 19.5)
               AND ( ( G.r - G.i - (G.g - G.r)/4 -0.18)
                      > (0.45 - 4*( G.g - G.r ) ) )
          AND ( (G.g - G.r ) >
                ( 1.35 + 0.25 *( G.r - G.i ) )
      ) ) )
```

**Q5.SXQL**

```
SELECT objID
FROM Galaxy
WHERE
  lDeV_r > 1.1 * lExp_r
  && lExp_r > 0
// Color cut for an elipical galaxy courtesy
// of James Annis of Fermilab
  && (objFlags & (OBJECT_BINNED1 | OBJECT_BINNED2
                          | OBJECT_BINNED4)) > 0
  && (objFlags & (OBJECT_BLENDED | OBJECT_NODEBLEND
            | OBJECT_CHILD)) != OBJECT_BLENDED
  && (objFlags & (OBJECT_EDGE | OBJECT_SATUR)) == 0
  && petroMag[3] > 17.5
  && (petroMag[2] > 15.5 || petroR50_r > 2)
  && (petroMag[2] > 0 && g>0 && r>0 && i>0)
  && ((petroMag[2]- reddening[2]) < 19.2
  && (petroMag[2] - reddening[2] < (13.1 +
       (7/3)*( g - reddening[1] - r + reddening[2] )
      +4 *(r-reddening[2]-i+reddening[3] )- 4*0.18 )
            )
  && ( ( r - reddening[2] - i + reddening[3] -
       ( g - reddening[1] - r + reddening[2])/4
         - 0.18 ) < 0.2 )
  && ( ( r - reddening[2] - i + reddening[3] -
       ( g - reddening[1] - r + reddening[2])/4
        -0.18 ) > -0.2 )  )
     ||(
        (petroMag[2] - reddening[2] < 19.5)
         &&( ( r - reddening[2] - i + reddening[3] -
             ( g - reddening[1] - r + reddening[2])/4
                - 0.18 ) >
                (0.45 - 4*( g - reddening[1] - r
                  + reddening[2] ) )
              )
         && ((g - reddening[1] - r + reddening[2]) >
               ( 1.35 + 0.25 *( r - reddening[2]
                             - i + reddening[3] ) )
             )   )
```

6. **Q8**: Find galaxies that are blended with a star and return the deblended galaxy magnitudes.

**Q8.SQL**

```
SELECT G.ObjID,G.u,G.g,G.r,G.i,G.z
FROM  Galaxy G, Star S
WHERE
  G.parentID > 0
  AND G.parentID = S.parentID
```

**Q8.SXQL**

```
SELECT objID,u,g,r,i,z
FROM Galaxy
WHERE
   EXIST(parent) &&
   parent.child{?}.objType == OBJECT_STAR
```

**7. Q9**: Find quasars with a line width >2000 km/s and 2.5<redshift<2.7.

**Q9.SQL**

```
DECLARE @qso int
SET @qso = dbo.fSpecClass('QSO')
DECLARE @hiZ_qso int
SET @hiZ_qso =dbo.fSpecClass('HIZ_QSO')
SELECT specObjID, z, zConf, SpecClass
FROM  SpecObj
WHERE
   ( SpecClass = @qso OR
     SpecClass = @hiZ_qso)
   AND  z BETWEEN 2.5 AND 2.7
   AND  zConf > 0.90
```

**Q9.SXQL**

```
SELECT spec_ID, z, zConf, specClass
FROM  SpecObj
WHERE
  ( specClass == SPEC_QSO ||
    specClass == SPEC_HIZ_QSO )
   AND  z BETWEEN 2.5 AND 2.7
   AND  zConf > 0.90
```

**8. Q10**: Find galaxies with spectra that have an equivalent width in Ha >40Å.

**Q10.SQL**

```
SELECT G.ObjID
FROM Galaxy   as G,
 SpecObj    as S,
 SpecLine   as L
WHERE G.ObjID = S.ObjID
  AND S.SpecObjID = L.SpecObjID
  AND L.LineId = 6565
  AND L.ew > 40
```

**Q10.SXQL**

```
SELECT  objID FROM (
   SELECT obj FROM (
      SELECT spec FROM SpecLine
      WHERE
         name.lineID == 6565 &&
         ew > 40 ) )
WHERE
  objType == OBJECT_GALAXY
```

**9. Q11**: Find all elliptical galaxies with spectra that have an anomalous emission line.

**Q11.SQL**

```
SELECT DISTINCT G.ObjID
FROM Galaxy as G,
  SpecObj  as S,
  SpecLine as L,
  XCRedshift as XC
WHERE G.ObjID = S.ObjID
  AND S.SpecObjID = L.SpecObjID
  AND S.SpecObjID = XC.SpecObjID
  AND XC.tempNo = 8
  AND L.lineID = 0
  AND L.ew > 10
  AND S.SpecObjID not in (
         SELECT S.SpecObjID
         FROM    SpecLine as L1
         WHERE S.SpecObjID= L1.SpecObjID
           AND abs(L.wave - L1.wave) <.01
           AND L1.LineID != 0
         )
```

**Q11.SXQL**

```
SELECT objID
FROM (
 SELECT obj FROM (
   SELECT spec FROM (
      SELECT found FROM SpecObj
      WHERE
         xcorrz{?}.tempNo == 8
   ) WHERE
     ew > 10
     && (((restWave - spec.measured{?}.restWave)
                   > -0.01)
     &&   ((restWave - spec.measured{?}.restWave)
                   < 0.01)) )
) WHERE
  objType == OBJECT_GALAXY
```

**10. Q15**: Provide a list of moving objects consistent with an asteroid.

**Q15.SQL**

```
SELECT objID,
  sqrt( power(rowv,2) + power(colv, 2) )
FROM PhotoObj
WHERE
  (power(rowv,2) + power(colv, 2)) > 50
   AND rowv >= 0 AND colv >=0
```

**Q15.SXQL**

```
SELECT objID,
   sqrt( rowv*rowv + colv*colv )
FROM PhotoObj
WHERE
  (rowv*rowv + colv*colv) > 50
   AND rowv >= 0 AND colv >=0
```

**11. Q16**: Search for Cataclysmic Variables and pre-CVs with White Dwarfs and very late secondaries.

| Q16.SQL | Q16.SXQL |
| --- | --- |
| SELECT run, camCol, rerun, field,<br> objID,u,g,r,i,z,<br> ra, dec,<br> count(*) as 'total',<br> sum(case when (type=3)<br>    then 1 else 0 end) as 'Galaxies',<br> sum(case when (type=6)<br>    then 1 else 0 end) as 'Stars',<br> sum(case when (type not in (3,6))<br>    then 1 else 0 end) as 'Other'<br>FROM PhotoPrimary<br>WHERE<br> u - g < 0.4 AND<br> g - r < 0.7 AND<br> r - i > 0.4 AND<br> i - z > 0.4 | SELECT RUN(), CAMCOL(), RERUN(), FIELDID(),<br>       OBJID(), ,g,r,i,z,<br>       RA(), DEC()<br>FROM Primary<br>WHERE<br>u - g < 0.4 &&<br>g - r < 0.7 &&<br>r - i > 0.4 &&<br>i - z > 0.4 |

**12. Q17**: Find all objects with velocities and errors (non-indexed quantities) in a given range.

| Q17.SQL | Q17.SXQL |
| --- | --- |
| SELECT  run, camCol, field, objID,<br> rowC,colC, rowV,colV,rowVErr,colVErr,<br> flags,<br> psfMag_u,psfMag_g,psfMag_r,psfMag_I,psfMag_z,<br> psfMagErr_u,psfMagErr_g,psfMagErr_r,<br> psfMagErr_I,psfMagErr_z<br>FROM PhotoPrimary<br>WHERE  ((rowv * rowv)/(rowvErr * rowvErr) +<br>       (colv * colv)/(colvErr * colvErr) > 4) | SELECT RUN(), CAMCOL(), RERUN(), FIELDID(),<br> OBJID(), rowC,colC, rowV,colV,rowVErr,colVErr,<br> objFlags,<br> psfMag,<br> psfMagErr<br>FROM PhotoPrimary<br>WHERE  ((rowv * rowv) / (rowvErr * rowvErr) +<br>       (colv * colv) / (colvErr * colvErr) > 4) |

**13. Q18**: Find all objects within a given coordinate cut (RA and Dec).

| Q18.SQL | Q18.SXQL |
| --- | --- |
| SELECT colc_g, colc_r<br>FROM  PhotoObj<br>WHERE (-0.642788 * cx + 0.766044 * cy>=0) AND<br>      (-0.984808 * cx - 0.173648 * cy <0) | SELECT obj.col[1], obj.col[2]<br>FROM  PhotoTag<br>WHERE (-0.642788 * cx + 0.766044 * cy>=0) AND<br>      (-0.984808 * cx - 0.173648 * cy <0) |

**14. Q19**: Search for objects and fields by their non-indexed short IDs.

| Q19.SQL | Q19.SXQL |
| --- | --- |
| SELECT  objID, field, ra, dec<br>FROM PhotoObj<br>WHERE obj = 14 AND  field = 11 | SELECT  objID, object, field.field, ra, dec<br>FROM (<br>      SELECT obj FROM field<br>       WHERE field == 11<br>      )<br>WHERE object == 14 |

15. **Q20**: Search for all galaxies with bluer centers: for all galaxies with Petrosian radius < 18 that are not saturated, not bright and not on the edge of the frame, give me those with centers appreciably bluer than their outer parts, i.e., define the center color as: u_psf - g_psf and define the outer color as: u_model - g_model; return all objects which have (u_model - g_model) - (u_psf - g_psf) < -0.4.

**Q20.SQL**

```
DECLARE @flags  BIGINT;
SET @flags =
dbo.fPhotoFlags('SATURATED') +
  dbo.fPhotoFlags('BRIGHT')    +
  dbo.fPhotoFlags('EDGE')
SELECT colc_u, colc_g,  objID
FROM  Galaxy
WHERE
  (Flags &  @flags )= 0
  AND petroRad_r < 18
  AND ((colc_u - colc_g) -
        (psfMag_u - psfMag_g)) < -0.4
```

**Q20.SXQL**

```
SELECT col[0], col[1],  objID
       obj.col[0], obj.col[1], objID
FROM Galaxy
WHERE
  (objFlags & (OBJECT_SATUR | OBJECT_BRIGHT |
OBJECT_EDGE) ) == 0
  && petroRad[2] < 18
  && ((obj.col[0] - obj.col[1]) -
       (psfMag[0] - psfMag[1])) < -0.4
```

16. **Q21**: Retrieve the PSF colors of all stars brighter than 20th magnitude, rejecting on various flags, that have PSP_STATUS = 2.

**Q21.SQL**

```
SELECT s.psfMag_g, s.run, s.camCol,
       s.rerun, s.field
FROM Star s, Field f
WHERE
  s.fieldid = f.fielded
  AND  s.psfMag_g < 20
  AND  f.pspStatus = 2
```

**Q21.SXQL**

```
SELECT  psfMag[1],RUN(), CAMCOL(),
        RERUN(), FIELDID()
FROM (  SELECT obj FROM Field
        WHERE
        pspStatus == 2 )
WHERE objType == OBJECT_STAR
  && ( (status & AR_OBJECT_STATUS_PRIMARY) > 0 )
  && psfMag[1] < 20
```

17. **Q22**: Cluster finding.

**Q22.SQL**

```
DECLARE @binned BIGINT
SET @binned =  dbo.fPhotoFlags('BINNED1')
            + dbo.fPhotoFlags('BINNED2')
            + dbo.fPhotoFlags('BINNED4')
DECLARE @deblendedChild BIGINT
SET @deblendedChild =
              dbo.fPhotoFlags('BLENDED')
            + dbo.fPhotoFlags('NODEBLEND')
            + dbo.fPhotoFlags('CHILD')
DECLARE @blended BIGINT
SET @blended = dbo.fPhotoFlags('BLENDED')
SELECT  camCol, run, rerun, field,
        objID, ra, dec
FROM Galaxy
WHERE (flags &  @binned )> 0
  AND (flags&@deblendedChild)!= @blended
  AND petroMag_i < 23
```

**Q22.SXQL**

```
SELECT
        CAMCOL(), RUN(), RERUN(), FIELDID(),
        OBJID(), RA(), DEC(),
        petroMag[3],
        objFlags
FROM Galaxy
WHERE
    ((objFlags & (OBJECT_BINNED1 | OBJECT_BINNED2 |
                       OBJECT_BINNED4)) > 0)
&& ((objFlags & (OBJECT_BLENDED | OBJECT_NODEBLEND |
                OBJECT_CHILD)) != OBJECT_BLENDED)
&&  petroMag[3] < 23
```

**18. Q23**: Diameter-limited sample of galaxies.

**Q23.SQL**

```
DECLARE @binned BIGINT
SET @binned =  dbo.fPhotoFlags('BINNED1') |
               dbo.fPhotoFlags('BINNED2') |
               dbo.fPhotoFlags('BINNED4')
DECLARE @deblendedChild BIGINT
SET @deblendedChild =
               dbo.fPhotoFlags('BLENDED') |
             dbo.fPhotoFlags('NODEBLEND') |
               dbo.fPhotoFlags('CHILD')
DECLARE @blended BIGINT
SET @blended = dbo.fPhotoFlags('BLENDED')
DECLARE @noPetro BIGINT
SET @noPetro = dbo.fPhotoFlags('NOPETRO')
DECLARE @tooLarge BIGINT
SET @tooLarge = dbo.fPhotoFlags('TOO_LARGE')
DECLARE @saturated BIGINT
SET @saturated =
                dbo.fPhotoFlags('SATURATED')
SELECT run,camCol,rerun,field,objID,ra,dec
FROM Galaxy
WHERE (flags &  @binned )> 0
  AND (flags& @deblendedChild)!= @blended
  AND (  (( flags & @noPetro = 0)
          AND petroRad_i > 15)
      OR ((flags & @noPetro > 0)
         AND petroRad_i > 7.5)
      OR ((flags & @tooLarge > 0)
         AND petroRad_i > 2.5)
      OR ((flags & @saturated  = 0 )
          AND petroRad_i > 17.5)
      )
```

**Q23.SXQL**

```
SELECT RUN(), CAMCOL(),  RERUN(), FIELDID(),
 OBJID(), RA(),  DEC()
FROM Galaxy WHERE (
  (objFlags & (OBJECT_BINNED1 | OBJECT_BINNED2 |
                  OBJECT_BINNED4)) > 0
&&(objFlags & (OBJECT_BLENDED | OBJECT_NODEBLEND |
                OBJECT_CHILD)) != OBJECT_BLENDED
&&  ( ((objFlags & OBJECT_NOPETRO == 0)
    && petroRad[3] > 15)
      ||((objFlags & OBJECT_NOPETRO > 0)
             && petroRad[3] > 7.5))
      ||  ( objFlags & OBJECT_TOO_LARGE > 0
             && petroRad[3] > 2.5 )
      ||  ( objFlags & OBJECT_SATUR == 0
             && petroRad[3] > 17.5 )
)
```

**19. Q24**: Search for extremely red galaxies.

**Q24.SQL**

```
DECLARE @binned BIGINT
SET @binned =  dbo.fPhotoFlags('BINNED1')
             + dbo.fPhotoFlags('BINNED2')
             + dbo.fPhotoFlags('BINNED4')
DECLARE @deblendedChild BIGINT
SET @deblendedChild =
           dbo.fPhotoFlags('BLENDED')
        +  dbo.fPhotoFlags('NODEBLEND')
        +    dbo.fPhotoFlags('CHILD')
DECLARE @blended BIGINT
SET @blended=dbo.fPhotoFlags('BLENDED')
DECLARE @crIntrp BIGINT
SET @crIntrp=dbo.fPhotoFlags('COSMIC_RAY')
             +   dbo.fPhotoFlags('INTERP')
SELECT g.run, g.camCol, g.rerun,
 g.field, g.objID, g.ra, g.dec
FROM Field f, Galaxy g
WHERE g.fieldid = f.fieldid
  AND (g.flags & @binned ) > 0
  AND (g.flags&@deblendedChild)!= @blended
  AND (g.flags &  @crIntrp ) = 0
  AND f.psfWidth_r < 1.5
  AND (g.i - g.z > 1.0)
```

**Q24.SXQL**

```
SELECT RUN(),RERUN(), CAMCOL(), FIELDID(),
 OBJID(), RA(), DEC()
FROM Galaxy
WHERE (
  ( (objFlags & (OBJECT_BINNED1 | OBJECT_BINNED2 |
      OBJECT_BINNED4)) > 0 )
&&  ( (objFlags & (OBJECT_BLENDED | OBJECT_NODEBLEND |
       OBJECT_CHILD)) != OBJECT_BLENDED )
&& ( (objFlags & (OBJECT_CR | OBJECT_INTERP)) == 0 )
&& field.psfWidth[2] < 1.5
&& ( I - z - (reddening[3] - reddening[4]) > 1.0 )
   )
```

**20. Q25**: The BRG (Bright Red Galaxy) sample.

| **Q25.SQL** | **Q25.SXQL** |
|---|---|
| ```DECLARE @binned BIGINT<br>SET @binned =  dbo.fPhotoFlags('BINNED1') \|<br>  dbo.fPhotoFlags('BINNED2') \|<br>  dbo.fPhotoFlags('BINNED4')<br>DECLARE @deblendedChild BIGINT<br>SET @deblendedChild =<br>dbo.fPhotoFlags('BLENDED')   \|<br>   dbo.fPhotoFlags('NODEBLEND') \|<br>   dbo.fPhotoFlags('CHILD')<br>DECLARE @blended BIGINT<br>SET @blended = dbo.fPhotoFlags('BLENDED')<br>DECLARE @edgedSaturated BIGINT<br>SET @edgedSaturated = dbo.fPhotoFlags('EDGE') \|<br>    dbo.fPhotoFlags('SATURATED')<br>SELECT<br> run, camCol, rerun, field, objID, ra, dec<br>FROM Galaxy<br>WHERE (<br>  ( flags &  @binned ) > 0<br>  AND ( flags &  @deblendedChild ) !=  @blended<br>  AND ( flags &  @edgedSaturated ) = 0<br>  AND petroMag_i > 17.5<br>  AND ( petroMag_r > 15.5 OR petroR50_r > 2 )<br>  AND ( petroMag_r > 0 AND g>0 AND r>0 AND i>0 )<br>  AND ( (petroMag_r-reddening_r) < 19.2<br>  AND ( petroMag_r - reddening_r <<br>   (13.1 + (7/3) * (g-r) + 4 * (r-i) - 4*0.18) )<br> AND ( (r - i - (g - r)/4 - 0.18) < 0.2 )<br>   AND ( (r - i - (g - r)/4 - 0.18) > -0.2 )<br>   AND ( (petroMag_r - reddening_r +  2.5 *<br>      LOG10(2 * 3.1415 * petroR50_r *<br>              petroR50_r)) < 24.2  ) )<br>   OR  ( (petroMag_r - reddening_r < 19.5)<br>     AND ( (r-i-(g-r)/4-0.18) > (0.45-4*(g-r))<br>     AND ( (g - r) > (1.35 + 0.25 * (r - i))) )<br>  AND ( (petroMag_r - reddening_r  +<br>        2.5 * LOG10( 2 * 3.1415 * petroR50_r *<br>petroR50_r )) < 23.3 ) )``` | ```SELECT<br> RUN(),CAMCOL(),RERUN(),FIELDID(),OBJID(),<br> RA(),DEC()<br>FROM Galaxy<br>WHERE (<br>  ( (objFlags & (OBJECT_BINNED1 \|<br>OBJECT_BINNED2 \| OBJECT_BINNED4)) > 0 ) &&<br>  ( (objFlags & (OBJECT_BLENDED \|<br>OBJECT_NODEBLEND \| OBJECT_CHILD)) !=<br>OBJECT_BLENDED ) &&<br>  ( (objFlags & (OBJECT_EDGE<br>        \| OBJECT_SATUR)) == 0 ) &&<br>  petroMag[3] > 17.65 &&<br>  ( petroMag[2] > 15.5 \|\| petroR50_r > 2 ) &&<br>  ( petroMag[2] > 0 && g>0 && r>0 && i>0 ) &&<br>  ( (petroMag[2] - reddening[2] < 19.2) &&<br>    (petroMag[2] - reddening[2] < 13.1 +<br>     (7/3)*(g-reddening[1]-r+reddening[2]) +<br>      4*(r - reddening[2] - i + reddening[3])-<br>      4 * 0.18) && ((r - reddening[2] - i +<br>        reddening[3] - (g - reddening[1] -<br>        r + reddening[2])/4 - 0.18) < 0.2) &&<br>     ((r - reddening[2] - i + reddening[3]<br>           - (g - reddening[1] -<br>     r + reddening[2])/4 - 0.18) > -0.2) &&<br>     ((petroMag[2] - reddening[2]) + 2.5*<br>     LOG(2*3.1415*petroR50_r*petroR50_r)<24.2)<br>  ) \|\|<br>  ( (petroMag[2] - reddening[2] < 19.5) &&<br>    ((r - reddening[2] - i + reddening[3] -<br>    (g-reddening[1]-r+reddening[2])/4-0.18) ><br>   0.45-4*(g-reddening[1]-r+reddening[2])) &&<br>    (g - reddening[1] - r + reddening[2]  ><br>     1.35 + 0.25 * (r - reddening[2] - i +<br>reddening[3])) &&<br>    ((petroMag[2]-reddening[2]) + 2.5 *<br>    LOG(2*3.1415*petroR50_r*petroR50_r)<23.3)<br>  ))``` |

**21. Q26**: Search for low redshift (z) QSO candidates.

| **Q26.SQL** | **Q26.SXQL** |
|---|---|
| ```SELECT g, run, rerun, camcol, field, objID<br>FROM Galaxy<br>WHERE (modelMag_g <= 22)<br>  AND (modelMag_u - modelMag_g >= -0.27)<br>  AND (modelMag_u - modelMag_g < 0.71)<br>  AND (modelMag_g - modelMag_r >= -0.24)<br>  AND (modelMag_g - modelMag_r < 0.35)<br>  AND (modelMag_r - modelMag_i >= -0.27)<br>  AND (modelMag_r - modelMag_i < 0.57)<br>    AND (modelMag_i - modelMag_z >= -0.35)<br>    AND (modelMag_i - modelMag_z < 0.70)``` | ```SELECT  g, RUN(), RERUN(), CAMCOL(),<br>        FIELDID(), OBJID()<br>FROM Galaxy<br>WHERE ( (g <= 22)<br>  && (u-g >= -0.27) && (u-g < 0.71)<br>  && (g-r >= -0.24) && (g-r < 0.35)<br>  && (r-i >= -0.27) && (r-i < 0.57)<br>  && (i-z >= -0.35) && (i-z < 0.70)<br>   )``` |

**22. Q27**: Check the errors on moving objects – compare the velocity to the error in velocity and see if the object is flagged as a moving object.

| Q27.SQL | Q27.SXQL |
|---|---|
| <pre>DECLARE @moved BIGINT<br>SET @moved = dbo.fPhotoFlags('MOVED')<br>DECLARE @badMove BIGINT<br>SET @badMove =<br>      dbo.fPhotoFlags('BAD_MOVING_FIT')<br>SELECT  run, rerun, camcol, field,<br>       objID, ra, dec,<br>       rowv, colv, rowvErr, colvErr,<br>       i,<br>       (flags & @moved) as MOVED,<br>       (flags & @badMove) as BAD_MOVING_FIT<br>FROM Galaxy<br>WHERE<br>  (flags & (@moved + @badMove)) > 0<br>  AND (rowv * rowv + colv * colv) >=<br> (rowvErr * rowvErr + colvErr * colvErr)</pre> | <pre>SELECT RUN(),RERUN(),CAMCOL(),FIELDID(),<br>       OBJID(), RA(),DEC(),<br>       obj.rowv, obj.colv, obj.rowvErr,<br>       obj.colvErr, i,<br>       objFlags & OBJECT_MOVED,<br>       objFlags & BAD_MOVING_FIT<br>FROM Galaxy<br>WHERE (<br>  ((objFlags & (OBJECT_MOVED \|<br>          OBJECT_BAD_MOVING_FIT)) > 0)<br>&& (((obj.rowv * obj.rowv) + (obj.colv * obj.colv))<br>>=   ((obj.rowvErr * obj.rowvErr)<br>        + (obj.colvErr * obj.colvErr)))<br>      )</pre> |

**23. Q28**: Extract a random sample of the data – get the colors of 100,000 random objects from all fields that are survey quality so that color-color diagrams can be made of them..

| Q28.SQL | Q28.SXQL |
|---|---|
| <pre>SELECT u,g,r,i,z<br>FROM Galaxy<br>WHERE<br>  (obj  %100 )= 1</pre> | <pre>SELECT u,g,r,i,z<br>FROM Galaxy<br>WHERE<br>  (obj.object - (obj.object/100) * 100) == 1</pre> |

**24. Q29**: Find quasars.

| Q29.SQL | Q29.SXQL |
|---|---|
| <pre>SELECT  run, camCol, rerun, field, objID,<br>        u,g,r,i,z,<br>        ra, dec<br>FROM Star      -- or sxGalaxy<br>WHERE ( modelMag_u - modelMag_g > 2.0<br>        OR u > 22.3 )<br>  AND ( modelMag_i < 19 )<br>  AND ( modelMag_i > 0 )<br>  AND ( modelMag_g - modelMag_r > 1.0 )<br>  AND ( modelMag_r - modelMag_i <<br>   (0.08+0.42*(modelMag_g - modelMag_r-0.96))<br>       OR modelMag_g - modelMag_r > 2.26 )<br> AND  ( modelMag_i - modelMag_z < 0.25 )</pre> | <pre>SELECT  RUN(), RERUN(),<br>        CAMCOL(), FIELDID(), OBJID(),<br>        u,g,r,i,z,<br>        RA(), DEC()<br>FROM Star<br>WHERE ( u - g > 2.0 OR u > 22.3 )<br>  AND ( i < 19 )<br>  AND ( i > 0 )<br>  AND ( g - r > 1.0 )<br>  AND ( r - i < (0.08 + 0.42 * (g - r - 0.96))<br>     OR g - r > 2.26 )<br> AND  ( i - z < 0.25 )</pre> |

**25. Q30**: Search for objects and fields by their non-indexed quantities.

| **Q30.SQL** | **Q30.SXQL** |
|---|---|
| <pre>SELECT g.run, g.rerun, g.camCol,<br>      f.field, p.objID, p.u,<br>      p.modelMagErr_u ,<br>      p.petroMag_r - p.reddening_r,<br>      p.petroMagErr_r,<br>      p.status & 0x00002000,<br>      f.psfWidth_r<br>FROM  photoobj p, field f, segment g  -- , tag t<br>WHERE<br>  f.fieldid = p.fieldid<br>  AND f.segmentid = g.segmentid<br>  AND f.psfWidth_r > 2<br>  AND p.colc > 1300.0</pre> | <pre>SELECT RUN(), RERUN(), CAMCOL(),<br>       FIELDID(), OBJID(),<br>       modelMag[0] - reddening[0],<br>       modelMagErr[0] ,<br>       petroMag[2] - reddening[2],<br>       petroMagErr[2],<br>       status & 0x00002000,<br>       field.psfWidth[2]<br>FROM ( SELECT obj<br>       FROM Field<br>       WHERE psfWidth[2] > 2<br>          && obj.colC > 1300.0<br>)</pre> |

**26. Q31**: Different version of Q30 with a much wider search (less limiting non-indexed constraint).

| **Q31.SQL** | **Q31.SXQL** |
|---|---|
| <pre>SELECT g.run, g.rerun, g.camCol, f.field,<br> p.objID, p.ra, p.dec, p.Rowc, p.Colc,<br> p.u, p.modelMagErr_u ,p.g, p.modelMagErr_g,<br> p.r, p.modelMagErr_r,<br> p.petroMag_r - p.reddening_r,<br> p.petroMagErr_r, p.i, p.modelMagErr_i, p.z,<br> p.status & 0x00002000, f.psfWidth_r<br>FROM<br> photoobj p, field f, segment g<br>WHERE<br>  f.fieldid = p.fieldid<br>  AND f.segmentid = g.segmentid<br>  AND p.colc > 400.0</pre> | <pre>SELECT  RUN(), RERUN(), CAMCOL(),  FIELDID(),<br>       OBJID(),  RA(), DEC(),<br>       rowC, colC,<br>       u - reddening[0], obj.modelMagErr[0] ,<br>       g - reddening[1], obj.modelMagErr[1],<br>       r - reddening[2], obj.modelMagErr[2],<br>       petroMag[2] - reddening[2],<br>       obj.petroMagErr[2],<br>       i - reddening[3], obj.modelMagErr[3],<br>       z - reddening[4],<br>       status & 0x00002000,<br>       field.psfWidth[2]<br>FROM PhotoTag<br>WHERE colC > 400.0</pre> |

**<u>Differences between SQL and SXQL query versions</u>**

1. The primary object tables in the MS-SQL version have de-reddened *u*,*g*,*r*,*i*,*z*, i.e. the correction for interstellar reddening of the light has been applied to the magnitudes recorded in the tables. However, the *u*,*g*,*r*,*i*,*z* magnitudes in the corresponding classes in the OODB have not been de-reddened. This makes most queries between the two not strictly equivalent. In order to make them equivalent, either the reddening correction is removed from the SQL version or added to the SXQL version as applicable.
2. Sometimes it was necessary to add the DISTINCT qualifier in the SELECT clause in the SQL version to avoid duplicates (e.g. Q11).
3. Some queries from the original set could not be translated to SXQL, for the reasons listed below for each of the queries. They highlight the limitations in the query language and data model that we implemented for the OODBMS.
   - Q6: Parent/child links were not present in object database, hence SQL queries that contained joins on the parented field could not be translated.
   - Q7: There is no GROUP BY clause or way to sort into buckets in SXQL.
   - Q12: It is not possible to do a grided count in SXQL.
   - Q13: There is no AVERAGE function or binning possible in SXQL.

- Q14: There is no pre-computed nearest neighbors list in the OODB.
- Q16: Aggregate functions could not be reproduced in SXQL.
- Q17-20: Same as Q14 – no nearest neighbors list in SXQL.